\documentclass[runningheads]{llncs}
\usepackage{graphicx}
\usepackage{amsmath}
\usepackage{amssymb}
\usepackage{textcomp, gensymb}
\usepackage{bbm}

\newcommand{\F}{\mathcal{F}}
\usepackage{hyperref}
\usepackage{xcolor}

\usepackage{amsmath,amssymb,stmaryrd}
\usepackage{multirow}

\begin{document}
%
\title{ICoNIK: Generating Respiratory-Resolved Abdominal MR Reconstructions Using Neural Implicit Representations in k-Space}
%
\titlerunning{ICoNIK}
%

\author{
Veronika Spieker \inst{1,2} 
\and
Wenqi Huang \inst{2}
\and
Hannah Eichhorn \inst{1,2}
\and
Jonathan Stelter \inst{3}
\and
Kilian Weiss \inst{4}
\and
Veronika A. Zimmer \inst{1,2}
\and
Rickmer F. Braren \inst{3,5}
\and
Dimitrios C. Karampinos \inst{3}
\and
Kerstin Hammernik \inst{2}
\and
Julia A. Schnabel \inst{1,2,6}
}

\institute{
Institute of Machine Learning in Biomedical Imaging, Helmholtz Munich, Germany,
\and School of Computation, Information and Technology, Technical University of Munich, Germany
\and School of Medicine, Technical University of Munich, Germany
\and Philips GmbH, Germany
\and German Cancer Consortium (DKTK), Partner Site Munich, Germany
\and School of Biomedical Engineering and Imaging Sciences, King’s College London, United Kingdom \\
\email{veronika.spieker@helmholtz-munich.de}}

\authorrunning{Spieker et al.}
%

\maketitle              
\begin{abstract}
Motion-resolved reconstruction for abdominal magnetic resonance imaging (MRI) remains a challenge due to the trade-off between residual motion blurring caused by discretized motion states and undersampling artefacts. In this work, we propose to generate blurring-free motion-resolved abdominal reconstructions by learning a neural implicit representation directly in k-space (NIK). Using measured sampling points and a data-derived respiratory navigator signal, we train a network to generate continuous signal values.  
To aid the regularization of sparsely sampled regions, we introduce an additional informed correction layer (ICo), which leverages information from neighboring regions to correct NIK's prediction. Our proposed generative reconstruction methods, NIK and ICoNIK, outperform standard motion-resolved reconstruction techniques and provide a promising solution to address motion artefacts in abdominal MRI. 


\end{abstract}
%
%
%

\section{Introduction}
Magnetic resonance imaging (MRI) is a non-invasive medical imaging modality with a high diagnostic value. However, its intrinsically long acquisition times make MRI more sensitive to motion than other imaging modalities. Especially respiration, which causes local non-rigid deformation of abdominal organs, induces non-negligible motion artefacts. Knowledge of the current state within the breathing cycle, e.g., using external or internal navigators, allows for selection of data to reconstruct from solely one breathing position and reduce artifacts \cite{McClelland_2013}. However, a large portion of data is discarded in this type of reconstruction, resulting in unnecessary prolongation of the acquisition time. 


Respiratory-resolved abdominal MRI reconstruction aims to provide high-quality images of one breathing position (typically at end-exhale), while leveraging acquired data points from all states in the respiratory cycle. To effectively utilize information from different breathing states, it is essential to be aware of the specific breathing state during data acquisition. Certain radial sampling trajectories, acquiring data points in a non-Cartesian manner, enable the derivation of a respiratory surrogate signal for motion navigation~\cite{Zaitsev_2015}.
Based on such a self-navigator, a common approach is to retrospectively bin the acquired data into multiple motion states and regularize over the motion states to obtain one high-quality reconstruction \cite{Feng_2016}. 
While a high number of motion states (i.e., high temporal resolution) decreases residual motion blurring, it minimizes the available data points per motion state. Consequently, undersampling artefacts occur due to the violation of the Nyquist criterion.

Deep learning has emerged as a powerful technique to cope with undersampling in MR, i.e., when fewer data points are available than required to reconstruct an image from the frequency domain \cite{Hammernik_2018,Hyun_2018}. Learned denoisers have shown promising results by leveraging information from multiple dynamics \cite{Jafari_2022,Kustner_2020,Terpstra_10.11.2022}, but require pretraining on fully sampled ground truth data. Deep generative models can be trained on acquired undersampled data to infer unavailable information, with the benefit of being independent of such expensive ground truth data \cite{Spieker_2023}. In particular for abdominal motion-resolved MR reconstruction, generative models have been proposed to infer a dynamic sequence of images from a latent space, either directly through a CNN \cite{Yoo_2021} or with intermediate motion modelling \cite{Zou_2022}. Lately,  Feng et al. \cite{Feng_31.12.2022} propose to learn a neural implicit representation of the dynamic image and regularize over multiple dynamics. While these approaches consider data consistency of the predicted images with the original acquired data in k-space, they all rely on binning of the dynamic data to generate images at some point of the training stage, risking residual motion blurring. Additionally, due to the the non-uniform sampling pattern, the non-uniform fast Fourier transform (NUFFT) \cite{Fessler_2007}, with computationally expensive operations such as regridding and density compensation, is required at each step of the optimization process. 
Recently, learning a neural implicit representation of k-space (NIK) \cite{Huang_2023} has shown promising results for binning-free ECG-gated cardiac reconstruction. The training and interpolation is conducted completely in the raw acquisition domain (k-space) and thereby, provide a way to avoid motion binning and expensive NUFFT operations within the optimization and at inference. However, in radial sampling patterns, sampling points sparsify when moving away from the k-space center to high frequency components, resulting in a compromised reconstruction of high frequency information.

In this work, we adapt NIK to respiratory-resolved abdominal MR reconstruction to overcome the challenge of motion-binning of classical reconstruction techniques. We extend NIK's capability to leverage classical parallel imaging concepts, i.e., that missing k-space points can be derived from neighboring points obtained with multiple coils \cite{Griswold_2002}. We perform an informed correction (ICo) by introducing a module which learns this multi-coil neighborhood relationship. Based on the inherently more densely sampled region in the center of k-space, we \textit{inform} the ICo module of the existing relationship by calibrating its weight and use it to \textit{correct} sparsely sampled high frequency regions. 
Our contributions are three-fold: 
\begin{enumerate}
    \item We learn the first neural implicit k-space representation (NIK) for motion-resolved abdominal reconstruction guided by a respiratory navigator signal.
    \item Inspired by classical MR reconstruction techniques, we perform an informed correction of NIK (ICoNIK), leveraging neighborhood information by applying a kernel which was auto-calibrated on a more densely sampled region.
    \item To demonstrate the potential of our work, we present quantitative and qualitative evaluation on retrospectively and prospectively undersampled abdominal MR acquisitions.
\end{enumerate}

\section{Methods}

\subsection{Motion Navigation and Classical Motion-Binned Reconstruction}
\label{sec:navigator}
\begin{figure}[h!]
    \centering
    \includegraphics[width=120mm]{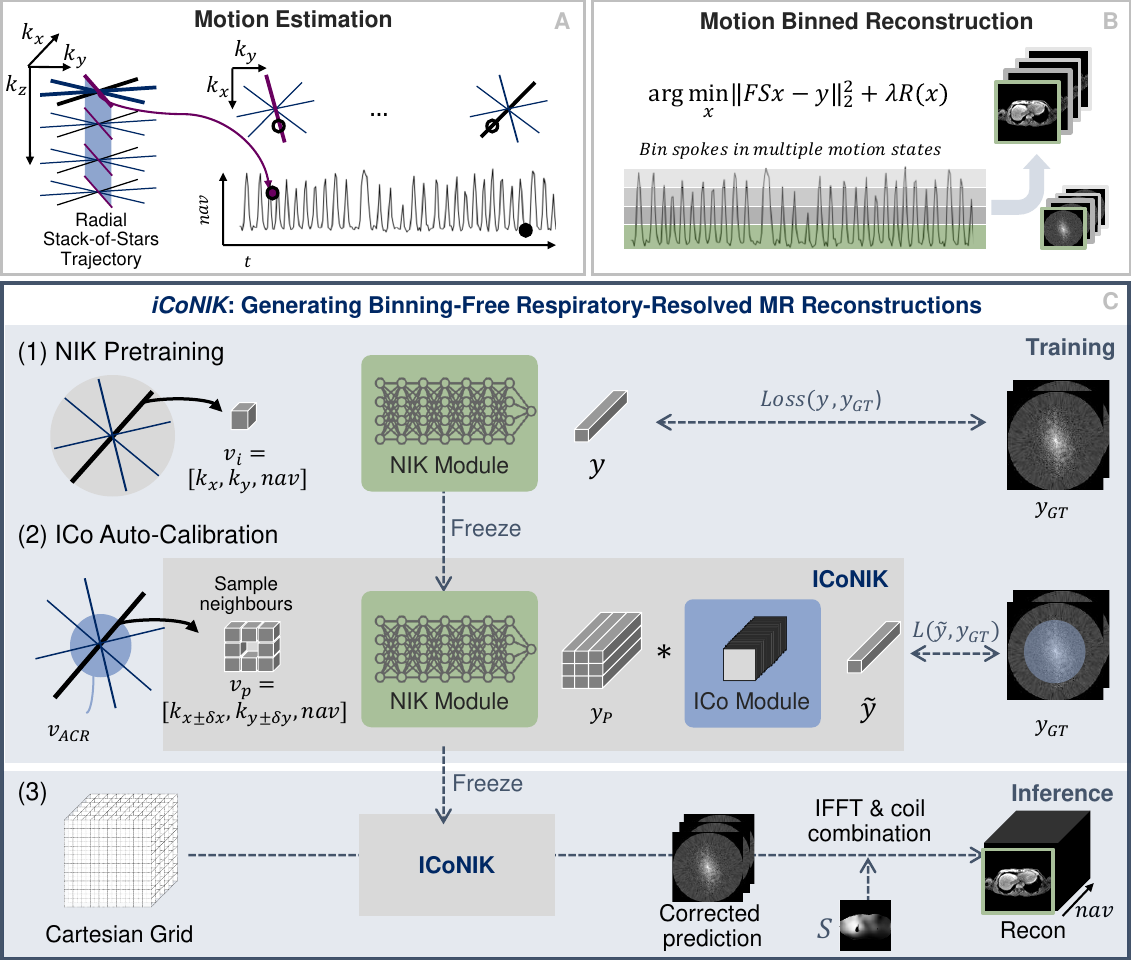}
    \caption{Graphical overview: (A) Derivation of a motion navigation signal $nav$ from the radial SoS trajectory, associating one $nav$ to each acquired $k_x/k_y$ sample. (B)~Classical motion-resolved reconstruction using motion binning based on the surrogate signal. (C) ICoNIK: (1) NIK is pretrained using samples of one slice obtained with the radial SoS and optimized with the corresponding measured values. Network weights are frozen for further processing. (2) Neighbours of sampling points are queried and their predicted values $p$ fused within the ICo layer. The ICo weights are optimized on sampled a restricted ACR. (3) At inference, a cartesian grid is sampled to generate the k-space and processed to generate dynamic reconstructions.}
    \label{fig:overview}
\end{figure}

In 3D abdominal imaging, knowledge of the current motion state can be deducted by acquiring data with a radial stack-of-stars (SoS) trajectory with Cartesian encoding in the feet-head direction. One spoke in the $k_x/k_y$ plane is acquired in each $k_z$ position before moving to the next partition, i.e., next set of spokes (Figure~\ref{fig:overview}A). By projecting the k-space center of each partition and applying principal component analysis, a 1D curve ($nav$) indicating the global relative feet-head motion over time $t$ can be derived. Since breathing motion is mainly driven in the feet-head direction, the extracted curve can be used as respiratory navigator signal. We refer the reader to \cite{Feng_2016} for more details on the respiratory navigator signal derivation.

Consecutively, the navigation signal is used in motion-resolved reconstruction methods to bin the spokes into a pre-defined number of dynamic states $n_d$ (Figure~\ref{fig:overview}B). A popular representative of motion-resolved reconstruction methods is XD-GRASP \cite{Feng_2016} which applies a total variation regularization in the dynamic motion state dimension. An inverse Fourier transform can be performed along this dimension and the dynamic images $x = x_{1...n_d}$ are obtained by solving the following optimization problem:

\begin{equation}
    \min_{x} \parallel \mathbf{\mathcal{F}}\mathbf{S}x-y \parallel_2^2 + \lambda \mathbf{\Phi}(x).
    \label{eq:GRASPrecon}
\end{equation}

where $y=y_{1...n_d}$ is the multi-coil radial k-space data sorted for each motion state $d$, $\mathcal{F}$ the NUFFT and $\mathbf{S}$ the coil sensitivity maps. Total variation in the temporal and spatial dimension is imposed by the second term, which consists of the finite difference operator $\mathbf{\Phi}$ and regularization weight $\lambda\in\mathbbm{R}^+$. 


\subsection{Binning-Free Neural Implicit k-Space Representation}
\label{sec:NIK}
To avoid binning of the acquired data, we propose to learn a continuous implicit representation of k-space (NIK) \cite{Huang_2023} conditioned on the data-derived respiratory navigator signal. The representation is learned by a neural network $G_\theta$ in the form of a mapping from k-space coordinates to the complex signal values. In contrast to the original NIK \cite{Huang_2023}, we predict the signal space for each coil simultaneously rather than including the coil dimension in the input to reduce training effort and enable further post-processing in the coil dimension.
Based on the radial sampling trajectory, $N$ coordinates $v_i=[nav_i, k_{xi},k_{yi}]$ $ i=1,2,\dots,N,$ $v_i \in\mathbbm{R}^{3}$ representing the current navigator signal value and trajectory position as well as its corresponding coil intensity values $y_{i} \in\mathbbm{R}^{n_c}$ can be queried (Fig. \ref{fig:overview}C.1).
The network $G_\theta$ is then optimized with the sampled data pairs $(v_i, y_{i})$ to approximate the underlying coordinate-signal mapping:
\begin{equation}
    \theta^* = \arg \min_\theta \left\|G_\theta(v) - y\right\|_2^2.
    \label{eq:dc_only}
\end{equation}
To account for the increased magnitude in the k-space center, the optimization is solved with a high-dynamic range loss, as described in \cite{Huang_2023}.
At inference, k-space values can be queried for any combination of coordinates within the range of the training data.
This allows for sampling based on a Cartesian grid $\bar{v}$ (Fig.~\ref{fig:overview}C), which enables a computationally efficient inverse Fourier transform $\F^{-1}$ instead of NUFFT. Final images are obtained by combining the inverse Fourier-transformed images using the complex conjugate coil sensitivities $\mathbf{S_c^H}$: 
\begin{equation}
    \hat{x} = \sum_{c}^{n_c} \mathbf{S_c^H} \F^{-1}(G_{\theta^*}(\bar{v})).
    \label{eq:inference}
\end{equation}

\subsection{Informed Correction of Neural Implicit k-Space (ICoNIK)}
\label{sec:PatchNIK}
The radial trajectory required for motion navigation and used to train NIK comes at the cost of increased data sparsity towards the outer edges of k-space, which represent the high frequencies and, therefore, details and noise in image domain. 
To increase the representation capability of NIK
, we include neighborhood information by processing NIK's multi-coil prediction with a informed correction module (ICo). Accelerated classical reconstruction methods have shown that k-space values can be derived by linearly combining the neighbouring k-space values of multiple coils. The set of weights can be auto-calibrated on a fully sampled region and consecutively applied to interpolate missing data points \cite{Griswold_2002,Seiberlich_2011}. We leverage this relationship and correct individual data points with the combination of the surrounding neighbors. 
As shown in Figure \ref{fig:overview}C.2, we sample $n_p$ neighboring points around the input coordinate $v_i$ to obtain $v_{p} \in\mathbbm{R}^{3 \times n_p}$. We predict its multi-coil signal values $y_{p} \in\mathbbm{R}^{n_c \times n_p}$ with NIK and fuse the information in the ICo module using a convolutional network $K_{\psi}$ to obtain the corrected signal values $\widetilde{y_{i}} \in\mathbbm{R}^{n_c}$ for $v_i$. 
\begin{equation}
    \widetilde{y_{i}} = K_{\psi} (G_{\theta^*}(v_{p})).
    \label{eq:inference}
\end{equation} 
In general, predictions in the center region of k-space are assumed to be more representative due to the higher amount of ground truth sampling points within this region (inherent by the radial trajectory). Therefore, we "inform" the correction module by calibrating it on NIK's predicted center of k-space, marked as autocalibration region (ACR) in Fig. \ref{fig:overview}C.2. We select sample points only within a certain distance $r$ to the k-space center for the optimization of the kernel weights $\psi^*$ and at inference, apply the informed correction module to regularize all predictions $y_i$ and output $\widetilde{y_{i}}$:

\begin{equation}
\label{eq:ACR}
    v_{ACR} \in V(d(v_i) < r), \text{ with } d(v_i) = \sqrt{k_x^2 + k_y^2},
\end{equation}
\begin{equation}
    \psi^* = \arg \min_\psi \left\|K_{\psi} (G_\theta^*(v_{ACR})) - y\right\|_2^2. 
    \label{eq:dc_only}
\end{equation}


\section{Experimental Setup}

\subsubsection*{Data}
Free-breathing golden angle stack-of-star acquisitions are obtained at 3T (Ingenia Elition X, Philips Healthcare) with a FOV=450$\times$450$\times$252~mm³, flip angle=10°, voxel size=1.5x1.5x3mm³, TR/TE1/TE2=4.9/1.4/2.7ms, $T_{shot}$=395ms after approval by the local ethics committee (approval number XXX). One prolonged sequence and one accelerated were acquired on separate subjects, resulting in datasets with 1800 (considered as acceleration R=1) and 600 radial spokes (R=3) with 600 samples per spoke ($n_{FE}$), respectively. Due to computational limitations and anisotropic spacing, further processing is conducted on 2D slices as proof-pf-principle. Coil sensitivity maps are estimated using ESPIRiT \cite{Uecker_2014}. 

\subsubsection*{Training and Inference}
We adapt NIK's architecture (Fourier encoding, 8 layers, 512 features) \cite{Huang_2023}
to output signal values in the coil dimension. We rescale the surrogate motion signal to [-1,1] and use the predefined selected number of spokes (1800 or 600) for training. NIK's training is stopped after 3000 epochs and the model with the lowest residual loss selected for further processing. The ICo module consists of three 3x3 complex kernel layers with interleaved complex ReLUs, and acts on neighboring samples spaced $\delta x, \delta y = n_{FE}^{-1}$ from the original coordinate. The ACR region was empirically determined as r = 0.4, and optimization is conducted for 500 epochs. Both modules are trained using an Adam optimizer with a learning rate of 3 $\cdot 10^{-5}$ and batch size 10000, optimizing for the linearized high dynamic range loss with $\sigma=1$, $\epsilon= 1 \cdot 10^{-2}$ and $\lambda=0.1$. Computations were performed on an NVIDIA RTX A6000, using Python 3.10.1 and PyTorch 1.13.1 (code available at: \url{https://github.com/vjspi/ICoNIK.git}). NIK/ICo module training took about 12/4 hours and 1.5/0.5 hours for the reference and the accelerated version, respectively. Reconstruction of 20 respiratory phases with a matrix size of 300x300 after training takes about 15 seconds.

\subsubsection*{Evaluation}
Reference motion-resolved 2D slices are reconstructed using inverse NUFFT (INUFFT) and XD-GRASP \cite{Feng_2016} (total variation in the spatial/temporal domain with factor 0.01/0.1) for 4 motion bins \cite{Feng_2016}. Results for the end-exhale state (first motion bin) are compared with the adapted NIK \cite{Huang_2023} and ICoNIK reconstructions at t within the same end-exhale state. For the prolonged sequence, the INUFFT of the gated data from one motion state is considered as approximately fully sampled and used as reference for computing quantitative evaluation measures, i.e., peak signal-to-noise ratio (PSNR), structural similarity index (SSIM) \cite{Wang_2004} and normalized root mean squared error (NRMSE) for 10 slices. Qualitative evaluation was performed on the full 1800 spokes (R1), retrospectively and prospectively downsampled 600 spokes (R3-retro/R3-pro). 

\section{Results}
The mean and standard deviation of the reconstruction results from 10 slices compared to the reference are shown in Table \ref{tab1}. Both generative binning-free methods, NIK and ICoNIK, significantly outperform XD-GRASP regarding SSIM, PSNR and NRMSE for both, the reference and accelerated acquisition. ICoNIK additionally shows increased PSNR and significantly reduced NRMSE for the reference scan R1 compared to NIK
For R1, a slight decrease of PSNR and increase NRMSE is noticable for ICoNIK compared to NIK.
\begin{table}[t]
\centering
\caption{Quantitative results for 10 slice reconstruction, (*) and ($^{+}$) mark statistical significance ($p<0.05$) compared to conventional XD-GRASP and NIK, respectively}
\label{tab1}
\begin{tabular}{l|ll|ll|ll}
\hline
\hline
& \multicolumn{2}{c}{SSIM $\uparrow$} & \multicolumn{2}{c}{PSNR (dB) $\uparrow$} & \multicolumn{2}{c}{NRMSE $\downarrow$} \\
\cline{2-7}
       & \multicolumn{1}{c}{R1} & \multicolumn{1}{c}{R3} & \multicolumn{1}{c}{R1} & \multicolumn{1}{c}{R3} & \multicolumn{1}{c}{R1} & \multicolumn{1}{c}{R3} \\
\hline
INUFFT & - & 0.80$\pm$0.01 & - & 26.55$\pm$0.68 & - & 0.65$\pm$0.03 \\
XD-GRASP & 0.85$\pm$0.01 & 0.79$\pm$0.01 & 28.19$\pm$0.73 & 25.73$\pm$0.56 & 0.56$\pm$0.03 & 0.73$\pm$0.03 \\
NIK & 0.91$\pm$0.01* & 0.82$\pm$0.01* & 32.01$\pm$0.89* & 27.47$\pm$0.75* & 0.35$\pm$0.02* & 0.59$\pm$0.03* \\
ICoNIK & 0.91$\pm$0.01* & 0.82$\pm$0.01* & 32.13$\pm$0.82* & 27.36$\pm$0.73* & 0.32$\pm$0.02*$^{+}$ & 0.60$\pm$0.02* \\
\hline
\hline
\end{tabular}
\end{table}



\begin{figure}[t]
    \centering
    \includegraphics[width=120mm]{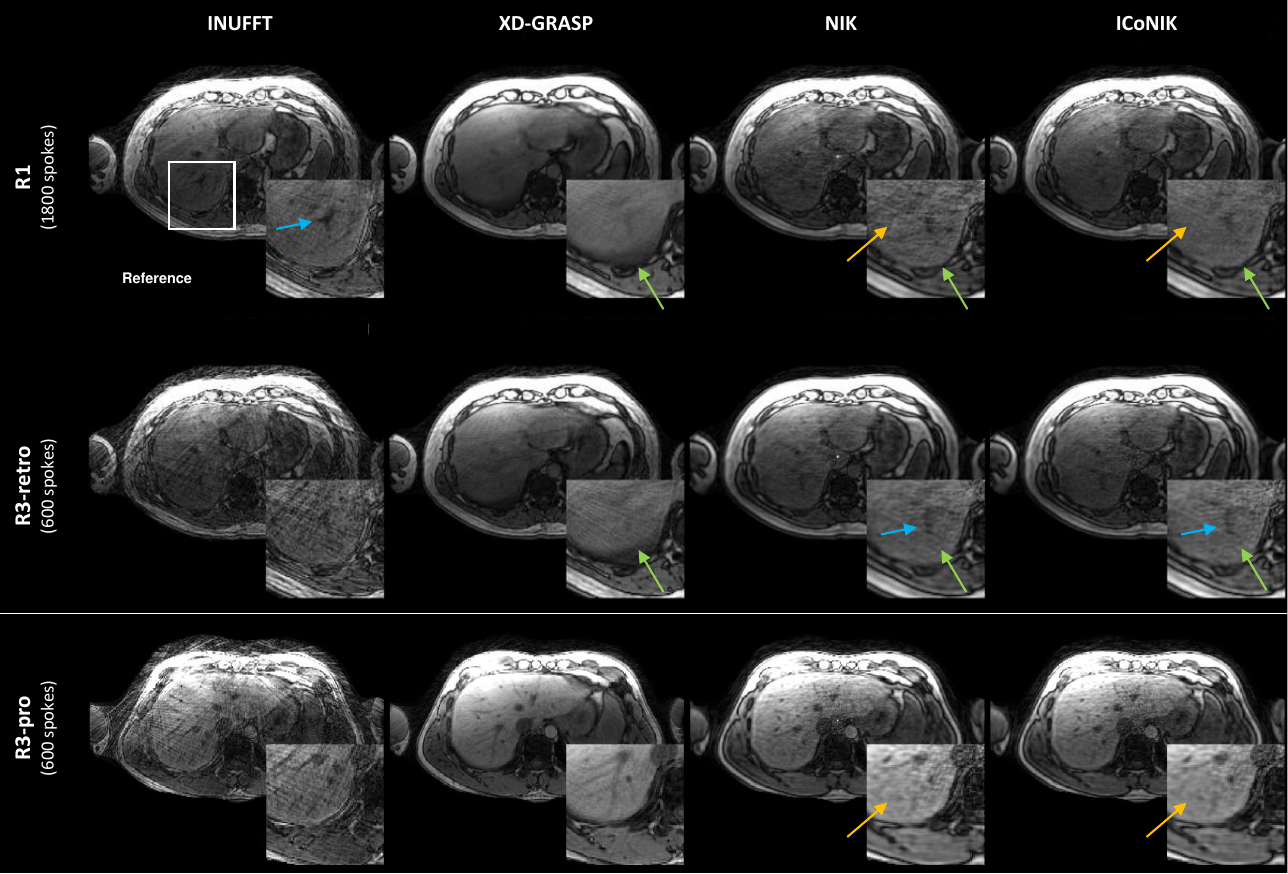}
    \caption{Motion-resolved reconstructions for the reference (R1), the retrospective downsampled (R3-retro) and  prospectively accelerated (R3-pro) acquisition. Both generative binning-free methods (NIK and ICoNIK) show less motion-blurring compared to traditional motion-binned reconstruction (green arrow) and maintain vessel structures (blue arrow). ICoNIK smooths the reconstruction results compared to NIK (orange arrow).}
    \label{fig:results}
\end{figure}

Qualitative reconstructions are visualized in Fig. \ref{fig:results}. Retrospective and prospective undersampling results (R3-retro and R3-pro) show the potential of XD-GRASP, NIK and ICoNIK to leverage information from different motion states to encounter undersampling artefacts originially present in INUFFT. The generative binning-free reconstruction methods (NIK and ICoNIK) indicate sharper images compared to the conventionally binned XD-GRASP, e.g., at the lung-liver edge (green arrow). Furthermore, vessel structures originally visible for the reference (R1-INUFFT) are only represented in NIK and ICoNIK (blue arrow). For the accelerated reconstructions (R3), undersampling artefacts are reduced for all motion-resolved methods and ICoNIK generates smoother reconstruction compared to NIK, while still maintaining structural information. This is supported by the quantitative finding of an increased PSNR for ICoNIK compared to NIK, while maintaining a similar SSIM. Since ICoNIK is capable of interpolating in the time dimension, an arbitrary number of motion states can be generated, allowing for a movie-like reconstruction (see suppl. material).

\section{Discussion and Conclusion}
In this work, we showed the potential of learning a neural implicit k-space representation to generate binning-free motion-resolved abdominal reconstructions. We leveraged parallel imaging concepts and induced neighborhood information within an informed correction layer to regularize the reconstruction. Due to the direct optimization in k-space and the learned dynamic representation based on the data-based respiratory navigator, we can generate motion-resolved images at any point of the breathing cycle. Both generative reconstruction methods, NIK and ICoNIK, outperform the traditional reconstruction approach. Additionally, ICoNIK is capable of smoothing reconstructions. While ICoNIK was developed for abdominal MR reconstruction using a radial SoS, we are confident that it can be transferred to other applications and sampling trajectories. 

The presented reconstruction technique benefits from its adaptability to non-uniform sampling patterns due to the continuous representation of k-space. Still, the errors may propagate when calibrating ICo on interpolated data, increasing the risk of erroneous calibration weights.
The improved performance of ICoNIK for R1 compared to R3, where less ground truth points are available for kernel calibration, supports this indication. Further investigation of the ideal patch retrieval method as well as ACR selection is planned. 

ICoNIK is inherently designed to handle inter-subject variability due to subject-specific reconstructions, but requires retraining for each application. Methods for preconditioned neural implicit networks are subject of further development to reduce the retraining cost. Similarly, information transfer between slices could be transferred to facilitate 3D applications at minimized computational cost. 
Intra-subject variation, e.g., due to changes in breathing pattern, can be captured by ICoNIK as long as the motion navigation signal is representative. Yet, bulk motion and temporal drifts may affect data-based motion surrogates, and thereby influence interpolation capability, motivating us to further look into the robustness of the neural implicit representation. 

Lastly, the difficulty to obtain a motion-free ground truth without binning data into motion states remains a significant obstacle in the evaluation process, also for our presented work. 
More reliable reference acquisition and evaluation techniques are still an active field of research, not only for learning-based reconstruction techniques based on deep generative models. To conclude, our promising findings encourage further investigation of combining traditional parallel imaging concepts with novel deep generative reconstruction models.


\section{Acknowledgements}
V.S. and H.E. are partially supported by the Helmholtz Association under the joint research school ”Munich School for Data Science - MUDS”.

%
%
%
\bibliographystyle{splncs04}
\newpage
\bibliography{references}




\end{document}